\documentclass[prb,preprintnumbers,amsmath,amssymb,floatfix]{revtex4}

\usepackage{graphicx}
\usepackage{subfigure}
\usepackage{dcolumn}

\begin{document}

\title{Full Quantum Theory of ${C_{60}}$ Double-slit Diffraction}
\author{Xiang-Yao Wu$^{a}$ \footnote{E-mail: wuxy2066@163.com},
Ji Ma$^{a}$, Bo-Jun Zhang$^{a}$, Hong Li$^{a}$, Xiao-Jing
Liu$^{a}$\\Nuo Ba$^{a}$, Si-Qi Zhang$^{a}$, Jing Wang$^{a}$, He
Dong$^{a}$ and Xin-Guo Yin$^{b}$ } \affiliation{a. Institute of
Physics, Jilin Normal University, Siping 136000 \\    b. Institute
of Physics, Huaibei Normal University, Huaibei 235000  }

\begin{abstract}

  In this paper, we apply the full new method of quantum theory to
study the double-slit diffraction of ${C_{60}}$ molecules. We
calculate the double-slit wave functions of ${C_{60}}$ molecules
by Schr\"{o}dinger equation, and calculate the diffraction wave
function behind the slits with the Feynman path integral quantum
theory, and then give the relation between the diffraction
intensity of double-slit and diffraction pattern position. We
compare the calculation results with two different double-slit
diffraction experiments. When the decoherence effects are
considered, the calculation
results are in good agreement with the two experimental data.\\
\vskip 5pt


\end{abstract}
\maketitle

\maketitle {\bf 1. Introduction} \vskip 8pt

Matter waves diffraction is a well established field in physics
which has become a large field of interest throughout the past
years [1, 2]. Diffractions with de-broglie waves have been
demonstrated for electrons and neutron, and extensively used for
fundamental tests of quantum-mechanical prediction [3-5] .
Recently, there are classical and quantum methods to study
interference and diffraction [6, 7]. As a matter of fact,
matter-wave interference and diffraction are quantum phenomena and
its full description needs quantum mechanical approach [8, 9]. At
present, Phenomena of diffraction have been studied in many
experiments [10-12]. Such as the electron diffraction experiment
in the crystal by Davisson and Germer in 1927. The electron single
and double slit diffraction experiment by J nsson in 1961. The
neutron single and double slit diffraction experiment by Anton
Zeilinger, Roland G hler and C.G.Shull in 1988, and these
experiments have been explained by some theoretical workers. In
view of quantum mechanics, the ${C_{60}}$ has the nature of wave,
and the wave is described by wave function $\psi(\vec{r},t)$,
which can be calculated with the Schr\"{o}dinger wave equation
[13, 14]. The wave function $\psi(\vec{r},t)$ has statistical
meaning, $\mid\psi(\vec{r},t)\mid^{2}$ can be explained as
particle's probability density at the definite position [15, 16].
In this paper, we apply the full new method of quantum theory to
study the double-slit diffraction of ${C_{60}}$ molecules, and the
${C_{60}}$ wave functions can be divided into three parts. The
first is the incoming area, the ${C_{60}}$ wave function is a
plane wave. The second is the slit area, where the ${C_{60}}$ wave
function can be calculated by the Schr\"{o}dinger wave equation.
The third is the diffraction area, where the ${C_{60}}$ wave
function can be obtained by Feynman path integral quantum theory.
Otherwise, we give the relation between the diffraction intensity
of double-slit and diffraction pattern position. When we consider
the decoherence effects in calculation, we find that the theory
results are in good agreement with the experimental data.

 \vskip 5pt
 \setlength{\unitlength}{0.1in}
\newpage
\begin{center}
\begin{figure}[tbp]
\includegraphics[width=7.5 cm]{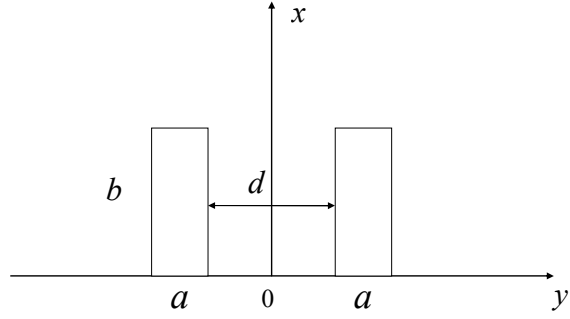}
\caption{${C_{60}}$ double-slit diffraction}
\end{figure}
\end{center}

{\bf 2. Quantum approach of ${C_{60}}$ diffraction}
 \vskip 8pt
 In an infinite plane, we consider a double-slit, its
width $a$, length $b$ and the slit-to-slit distance $d$ are shown
in FIG. 1. The $x$ axis is along the slit length $b$ and the $y$
axis is along the slit width $a$. We calculate the ${C_{60}}$ wave
function in the left slit with the Schr\"odinger equation, and the
${C_{60}}$ wave function of the right slit can be obtained easily.
At time $t$, we suppose that the incoming plane wave travels along
the $z$ axis. It is
\begin{equation}
\vec{\psi}_{0}(z, t)=\vec{A}\exp{\frac{i}{\hbar}(pz-Et)},
\end{equation}
where $A$ is a constant. The time-dependent Schr\"odinger equation
is
\begin{equation}
[-\frac{\hbar^{2}}{2M}\nabla^{2}+V(r)]{\psi}(\vec{r},
t)=E{\psi}(\vec{r}, t),
\end{equation}
where $M(E)$ is the mass(energy) of the ${C_{60}}$. The potential
in the left slit is
\begin{eqnarray}
V(x,y,z)= \left \{ \begin{array}{ll}
   0  \hspace{0.3in} 0\leq x\leq b, -\frac{d}{2}-a\leq y\leq -\frac{d}{2}, 0\leq z\leq c, \\
   \infty  \hspace{0.3in}  otherwise,
   \end{array}
   \right.
\end{eqnarray}
where $c$ is the thickness of slit.

When $V=0$, the time-independent Schr\"odinger equation in left
slit is
\begin{equation}
-\frac{\hbar^{2}}{2M}\nabla^{2}{\psi_1}(\vec{r})=E{\psi_1}(\vec{r}),
\end{equation}
the partial differential Eq. (4) can be solved by the method of
separation of variable.
\begin{equation}
\frac{\partial^{2}\psi_1(\vec{r})}{\partial
x^{2}}+\frac{\partial^{2}\psi_1(\vec{r})}{\partial
y^{2}}+\frac{\partial^{2}\psi_1(\vec{r})}{\partial
z^{2}}+\frac{2ME}{\hbar^{2}}\psi_1(\vec{r})=0,
\end{equation}
the wave function ${\psi_1}(x,y,z)$ satisfies the boundary
conditions
\begin{equation}
{\psi_1}(0,y,z)={\psi_1}(b,y,z)=0,\nonumber
\end{equation}
\begin{equation}
{\psi_1}(x,-a-\frac{d}{2},z)={\psi_1}(x,-\frac{d}{2},z)=0.
\end{equation}
By the method of separation of variable
${\psi_1}(x,y,z)=X(x)Y(y)Z(z)$, the general solution of Eq. (4) is
\begin{eqnarray}
\psi_{1mn} (x,y,z)&=&\sum_{m,n}(D_{mn}\sin{\frac{n\pi
x}{b}}\cos{\frac{m\pi y}{a}}\nonumber\\&& +D'_{mn}\sin{\frac{n\pi
x}{b}}\sin{\frac{m\pi y}{a}})
  \exp{[i\sqrt{\frac{2ME}{\hbar^{2}}-\frac{n^{2}\pi^{2}}{b^{2}}-\frac{m^{2}\pi^{2}}{a^{2}}}z]},
\end{eqnarray}
the general solution of time-dependent Schr\"odinger equation is
\begin{eqnarray}
\psi_{1mn}(x,y,z,t)&=&\sum_{m,n}(D_{mn}\sin{\frac{n\pi
x}{b}}\cos{\frac{m\pi y}{a}}\nonumber\\&& +D'_{mn}\sin{\frac{n\pi
x}{b}}\sin{\frac{m\pi y}{a}})
  \exp{[i\sqrt{\frac{2ME}{\hbar^{2}}-\frac{n^{2}\pi^{2}}{b^{2}}-\frac{m^{2}\pi^{2}}{a^{2}}}z]}e^{-\frac{i}{\hbar}Et},
\end{eqnarray}
since the wave functions are continuous at $z=0$, we have
\begin{equation}
\vec{\psi}_{0}(x,y,z,t)\mid_{z=0}=\vec{\psi}_{1mn}(x,y,z,t)\mid_{z=0}.
\end{equation}
Then
\begin{eqnarray}
A_1=\sum_{m,n}(D_{mn}\sin{\frac{n\pi x}{b}}\cos{\frac{m\pi
y}{a}}+D'_{mn}\sin{\frac{n\pi x}{b}}\sin{\frac{m\pi y}{a}}),
\end{eqnarray}
we can obtain the Fourier coefficient $D_{mn}$ and $D'_{mn}$ by
Fourier transform
\begin{eqnarray}
D_{mn}&=&\frac{4}{a
b}\int^{-\frac{d}{2}}_{-\frac{d}{2}-a}\int^{b}_{0}A\sin{\frac{n\pi
x}{b}}\cos{\frac{m\pi y}{a}}dx dy \nonumber\\
&=&-\frac{16A}{(2m+1)(2n+1)\pi^{2}}\sin{\frac{(2m+1)\pi
}{2a}}d\hspace{0.6in}\ m,n=0,1,2, 3,
\end{eqnarray}
\begin{eqnarray}
D'_{mn}&=&\frac{4}{a
b}\int^{-\frac{d}{2}}_{-\frac{d}{2}-a}\int^{b}_{0}A\sin{\frac{n\pi
x}{b}}\sin{\frac{m\pi y}{a}}dx dy \nonumber\\
&=&-\frac{16A}{(2m+1)(2n+1)\pi^{2}}\cos{\frac{(2m+1)\pi
}{2a}}d\hspace{0.6in}\ m,n=0,1,2, 3,
\end{eqnarray}
substituting (11) and (12) into (8), we can obtain the ${C_{60}}$
wave function in the left slit,

\begin{eqnarray}
\psi_{1}(x,y,z,t)&=&-\sum_{m,n}\frac{16A_1}{(2m+1)(2n+1)\pi^{2}}\exp{[i\sqrt{\frac{2ME}{\hbar^{2}}-\frac{(2n+1)^{2}\pi^{2}}{b^{2}}
-\frac{(2m+1)^{2}\pi^{2}}{a^{2}}}z]}\exp{[-\frac{i}{\hbar}Et]}
\nonumber\\&& \cdot
[\sin{\frac{(2m+1)\pi}{2a}d}\sin{\frac{(2n+1)\pi
x}{b}}\cos{\frac{(2m+1)\pi
y}{a}}\nonumber\\&&+\cos{\frac{(2m+1)\pi
}{2a}d}\sin{\frac{(2n+1)\pi x}{b}}\sin{\frac{(2m+1)\pi y}{a}}].
\end{eqnarray}

Similarly, we can obtain the ${C_{60}}$ wave function in the right
slit

\begin{eqnarray}
\psi_{2}(x,y,z,t)&=&-\sum_{m,n}\frac{16A_2}{(2m+1)(2n+1)\pi^{2}}\exp{[i\sqrt{\frac{2ME}{\hbar^{2}}-\frac{(2n+1)^{2}\pi^{2}}{b^{2}}
-\frac{(2m+1)^{2}\pi^{2}}{a^{2}}}z]}\exp{[-\frac{i}{\hbar}Et]}
\nonumber\\&& \cdot
[\sin{\frac{(2m+1)\pi}{2a}d}\sin{\frac{(2n+1)\pi
x}{b}}\cos{\frac{(2m+1)\pi
y}{a}}\nonumber\\&&-\cos{\frac{(2m+1)\pi
}{2a}d}\sin{\frac{(2n+1)\pi x}{b}}\sin{\frac{(2m+1)\pi y}{a}}].
\end{eqnarray}
\vskip 8pt

\begin{center}
\begin{figure}[tbp]
\includegraphics[width=6.0 cm]{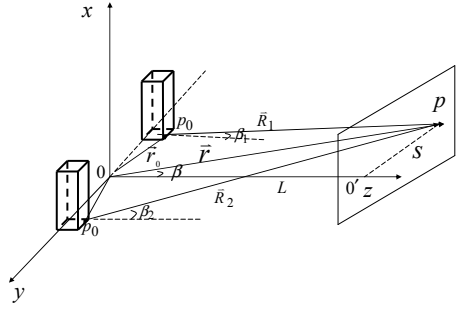}

\caption{${C_{60}}$ diffraction area diagram}
\end{figure}
\end{center}
\newpage
{\bf 3. The wave function of ${C_{60}}$ diffraction} \vskip 8pt
With Path Integral approach, we can calculate the ${C_{60}}$ wave
function in the diffraction area. ${C_{60}}$ diffraction area
diagram is shown in FIG. 2. $p_{0}$ is the position of a point on
the surface (z=c), its position vector is $\vec{r}_0$, $p$ is an
arbitrary point in the diffraction area and its position vector is
$\vec{r}$. At time $t_{0}$, the particles at position $p_{0}$. By
Eq. (13), we have
\begin{eqnarray}
\psi_{1}(r_{0},t_{0})&=&-\sum_{m,n}\frac{16A_1}{(2m+1)(2n+1)\pi^{2}}\exp{[i\sqrt{\frac{2ME}{\hbar^{2}}-\frac{(2n+1)^{2}\pi^{2}}{b^{2}}
-\frac{(2m+1)^{2}\pi^{2}}{a^{2}}}c]}\exp{[-\frac{i}{\hbar}Et]}
\nonumber\\&& \cdot
[\sin{\frac{(2m+1)\pi}{2a}d}\sin{\frac{(2n+1)\pi
x_{0}}{b}}\cos{\frac{(2m+1)\pi
y_{0}}{a}}\nonumber\\&&+\cos{\frac{(2m+1)\pi
}{2a}d}\sin{\frac{(2n+1)\pi x_{0}}{b}}\sin{\frac{(2m+1)\pi
y_{0}}{a}}].
\end{eqnarray}
The diffraction wave function can be calculated by Path Integral
formula[11], it is
\begin{equation}
\psi_{1}(\vec{r},t)=\int\\k(\vec{r},t;\vec{r_{0}},t_{0})\psi_{1}(r_{0},t_{0})dr_{0},
\end{equation}
the propagator $k(\vec{r_1},t;\vec{r_{0}},t_{0})$ is
\begin{equation}
k(\vec{r},t;\vec{r_{0}},t_{0})=[\frac{M}{2\pi
i\hbar(t-t_{0})}]^{\frac{3}{2}}\cdot
\exp{[\frac{iMR_1^{2}}{2\hbar(t-t_{0})}]},
\end{equation}
where $R$ is the distance between $p_{0}$ and $p$, and $(t-t_{0})$
is the time of ${C_{60}}$ propagating from $p_{0}$ to $p$.

Substituting Eq. (17) into Eq. (16),we have
\begin{equation}
\psi_{1}( \vec{r},t)=\int[\\\frac{M}{2\pi
i\hbar(t-t_{0})}]^{\frac{3}{2}}\cdot
\exp[\frac{iMR_1^{2}}{2\hbar(t-t_{0})}]\psi_{0}(r_{0},t_{0})dr_{0},
\end{equation}
where $dr_{0}=dx_{0}dy_{0}$.
\\
From Fig. 2, there is
\begin{eqnarray}
R_1^{2}&=&|\vec{r}-\vec{r_{0}}|^{2}=(x-x_{0})^{2}+(y-y_{0})^{2}+(z-c)^{2}\nonumber\\&\approx&
x^{2}+y^{2}+z^{2}-2xx_{0}-2yy_{0}-2zc+x_{0}^{2}+y_{0}^{2}+c^{2}\
\end{eqnarray}
In Eq. (19), $x_{0}^{2}$, $y_{0}^{2}$ and $c^{2}$ are second-order
infinitesimal, they can be ignored, then we have
\begin{equation}
R_1^{2}=r^{2}-2r\sin\alpha\cdot x_{0}-2r\sin\beta_1\cdot
y_{0}-2r\cos\theta\cdot c,
\end{equation}

where $\alpha$ is the angle between $r$ and $yz$ plane, $\beta_1$
is the angle between $R_1$ and $xz$ plane, $\theta$ is the angle
between $r$ and $z$ axis.

Substituting Eq. (15) and (20) into Eq. (18), the diffraction wave
function of the left slit is
\begin{eqnarray}
\psi_{out1}(\vec{r},t)&=&[\frac{M}{2\pi
i\hbar(t-t_{0})}]^\frac{3}{2}\exp{[\frac{iMr^{2}}{2\hbar(t-t_{0})}]}\exp{({-\frac{i}{\hbar}Et_{0}})}\exp[-\frac{iMr\cos\theta}{\hbar(t-t_{0})}\cdot
c]\nonumber\\&&
\sum_{m=0}^{\infty}\sum_{n=0}^{\infty}\frac{-16A_1}{(2m+1)(2n+1)\pi^2}
\exp{[i\sqrt{\frac{2ME}{\hbar^{2}}-(\frac{(2n+1)\pi}{b})^{2}-(\frac{(2m+1)\pi}{a})^{2}}\cdot
c]}\nonumber\\&&
\int^{b}_{0}\exp{[-\frac{iMr\sin\alpha}{\hbar(t-t_{0})}\cdot
x_{0}]}\sin\frac{(2n+1)\pi}{b}x_{0}dx_{0}[\sin\frac{(2m+1)\pi
d}{2a}\cdot\int^{-\frac{d}{2}}_{-\frac{d}{2}-a}\exp{[-\frac{iMr\sin\beta_{1}}{\hbar(t-t_{0})}\cdot
y_{0}]}\nonumber\\&&\cos\frac{(2m+1)\pi}{a}y_{0}dy_{0}+\cos\frac{(2m+1)\pi
d}{2a}\int^{-\frac{d}{2}}_{-\frac{d}{2}-a}\exp{[-\frac{iMr\sin\beta_{1}}{\hbar(t-t_{0})}\cdot
y_{0}]}\sin\frac{(2m+1)\pi}{a}y_{0}dy_{0}].
\end{eqnarray}

With the De Broglie relation $p=\frac{h}{\lambda}$, we have
\begin{equation}
k=\frac{2\pi}{\lambda}=\frac{Mv}{\hbar},
\end{equation}
i.e.,
\begin{equation}
k=\frac{MR_1}{\hbar(t-t_{0})},
\end{equation}
so
\begin{equation}
[\frac{M}{2\pi i\hbar(t-t_{0})}]^\frac{3}{2}=[\frac{MR_1}{2\pi
i\hbar(t-t_{0})R_1}]^\frac{3}{2}=(\frac{k}{2\pi
i\hbar})^{\frac{3}{2}}=(-\frac{\sqrt{2}}{2}-\frac{\sqrt{2}}{2}i)(\frac{k}{2\pi
r})^{\frac{3}{2}},
\end{equation}
and
\begin{equation}
\exp{[\frac{iMr^{2}}{2\hbar(t-t_{0})}]}=\exp{(\frac{ikr}{2})}.
\end{equation}
Substituting Eqs. (23)-(25) into Eq. (21), the diffraction wave
function of the left slit is
\begin{eqnarray}
\psi_{out1}(\vec{r},t)&=&(-\frac{\sqrt{2}}{2}-\frac{\sqrt{2}}{2}i)(\frac{k}{2\pi
r})^{\frac{3}{2}}\exp{(\frac{ikr}{2})}\exp{({-\frac{i}{\hbar}Et_{0}})}\exp[-ik\cos\theta\cdot
c]\nonumber\\&&
\sum_{m=0}^{\infty}\sum_{n=0}^{\infty}\frac{-16A_1}{(2m+1)(2n+1)\pi^2}
\exp{i\sqrt{\frac{2ME}{\hbar^{2}}-(\frac{(2n+1)\pi}{b})^{2}-(\frac{(2m+1)\pi}{a})^{2}}\cdot
c}\nonumber\\&& \int^{b}_{0}\exp{[-ik\sin\alpha\cdot
x_{0}]}\sin\frac{(2n+1)\pi}{b}x_{0}dx_{0}[\sin\frac{(2m+1)\pi
d}{2a}\cdot\int^{-\frac{d}{2}}_{-\frac{d}{2}-a}\exp{[-ik\sin\beta_{1}\cdot
y_{0}]}\nonumber\\&&\cos\frac{(2m+1)\pi}{a}y_{0}dy_{0}+\cos\frac{(2m+1)\pi
d}{2a}\int^{-\frac{d}{2}}_{-\frac{d}{2}-a}\exp{[-ik\sin\beta_{1}\cdot
y_{0}]}\sin\frac{(2m+1)\pi}{a}y_{0}dy_{0}],
\end{eqnarray}
Similarly, the diffraction wave function of the right slit is
\begin{eqnarray}
\psi_{out2}(\vec{r},t)&=&(-\frac{\sqrt{2}}{2}-\frac{\sqrt{2}}{2}i)(\frac{k}{2\pi
r})^{\frac{3}{2}}\exp{(\frac{ikr}{2})}\exp{({-\frac{i}{\hbar}Et_{0}})}\exp[-ik\cos\theta\cdot
c]\nonumber\\&&
\sum_{m=0}^{\infty}\sum_{n=0}^{\infty}\frac{-16A_2}{(2m+1)(2n+1)\pi^2}
\exp{[i\sqrt{\frac{2ME}{\hbar^{2}}-(\frac{(2n+1)\pi}{b})^{2}-(\frac{(2m+1)\pi}{a})^{2}}\cdot
c]}\nonumber\\&& \int^{b}_{0}\exp{[-ik\sin\alpha\cdot
x_{0}]}\sin\frac{(2n+1)\pi}{b}x_{0}dx_{0}[\sin\frac{(2m+1)\pi
d}{2a}\cdot\int^{\frac{d}{2}+a}_{\frac{d}{2}}\exp{[-ik\sin\beta_{2}\cdot
y_{0}]}\nonumber\\&&\cos\frac{(2m+1)\pi}{a}y_{0}dy_{0}-\cos\frac{(2m+1)\pi
d}{2a}\int^{\frac{d}{2}+a}_{\frac{d}{2}}\exp{[-ik\sin\beta_{2}\cdot
y_{0}]}\sin\frac{(2m+1)\pi}{a}y_{0}dy_{0}].
\end{eqnarray}
Where $\beta_2$ is the angle between $R_2$ and $xz$ plane, from
FIG. 2, since $a+d\ll L$, there is
$\beta_1\simeq\beta_2\approx\beta$ and $\theta$, $\alpha$ and
$\beta$ are satisfied the relation:
$\cos^{2}\theta+\sin^{2}\alpha+\sin^{2}\beta=1$.

The total diffraction wave function for the double-slit is
\begin{eqnarray}
\psi(x,y,z,t)=c_{1}\psi_{out1}(x,y,z,t)+c_{2}\psi_{out2}(x,y,z,t),
\end{eqnarray}

where $c_{1}$,$c_{2}$ satisfy the equation
\begin{equation}
c^{2}_{1}+c^{2}_{2}=1.
\end{equation}
For the double-slit diffraction, we can obtain the relative
diffraction intensity $I$ on the display screen
\begin{eqnarray}
I\propto|\vec{\psi}(x,y,x,t)|^{2}=c^{2}_{1}|\psi_{out1}(x,y,z,t)|^{2}+c^{2}_{2}|\psi_{out2}(x,y,z,t)|^{2}\nonumber\\
+2c_{1}c_{2}Re[\psi^{*}_{out1}(x,y,z,t)\psi_{out2}(x,y,z,t)].
\end{eqnarray}
\vskip 8pt {\bf 4. The relative diffraction intensity $I$ on the
display screen} \vskip 8pt Decoherence is introduced here using a
simple phenomenological theoretical model that assumes an
exponential damping of the interferences [17], i.e., the
decoherence is the dynamic suppression of the interference terms
owing to the interaction between system and environment. Eq. (28)
describes the coherence state coherence superposition, without
considering the interaction of system with external environment.
When we consider the effect of external environment, the total
wave function of system and environment for the double-slit
factorizes as [17]
\begin{eqnarray}
\psi_{out}(x,y,z,t)=c_{1}\psi_{out_{1}}(x,y,z,t)\otimes|E_{1}>_{t}+c_{2}\psi_{out_{2}}(x,y,z,t)\otimes|E_{2}>_{t},
\end{eqnarray}
where $\otimes|E_{1}>_{t}$ and $\otimes|E_{2}>_{t}$ describe the
state of the environment. Now, the diffraction intensity on the
screen is given by [3]
\begin{equation}
I=(1+|\alpha_{t}|^{2})[c_{1}^{2}|\psi_{out1}(\vec{r},t)|^{2}+c_{2}^{2}|\psi_{out2}(\psi_{out1}
(\vec{r},t)|^{2}+2c_{1}c_{2}\Lambda_{t}R{e}(\psi_{out1}^{*}(\vec{r},t)
\psi_{out2}(\vec{r},t))],
\end{equation}
where $\alpha_{t}=_{t}<E_{2}|E_{1}>_{t}$, and
$\Lambda_{t}=\frac{2|\alpha_{t}|^{2}}{1+|\alpha_{t}|^{2}}$. Thus,
$\Lambda_{t}$ is defined as the quantum coherence degree. The
fringe visibility of n is defined as [17]
\begin{equation}
v=\frac{I_{max}-I_{min}}{I_{max}+I_{min}},
\end{equation}
where $I_{max}$ and $I_{min}$ are the intensities corresponding to
the central maximum and the first minimum next to it,
respectively. The value for the fringe visibility of $\nu=0.625$
is obtained in the experiment [18], and the quantum coherence
degree $v\simeq\Lambda_{t}$ [17]. Eq. (32) is the diffraction
intensity of ${C_{60}}$ double-slit diffraction including
decoherence effects, and Eq. (30) is the diffraction intensity of
 ${C_{60}}$ double-slit diffraction considering coherence superposition.

 \vskip 8pt {\bf
 5. Numerical result} \vskip 8pt

 Next, we shall give out the calculation results of ${C_{60}}$
double-slit diffraction intensity $I$, and compare the calculation
results with experiment data, which were carried out by two
diffraction experiment device with different experiment parameters
and experiment data in Ref. [18] and Ref. [19]. In Ref. [18] and
Ref. [19], the authors have given the relation between diffraction
intensity and pattern position. In Eqs. (26) and (27), we should
convert diffraction angle $\beta$ to positions from FIG. 2, we can
find the relation is
$\sin\beta=\frac{s}{R}=\frac{s}{\sqrt{L^{2}+s^{2}}}$, where
$R=\mid\vec{r}\mid$ and $L$ is the distance between the slit and
screen, which are shown in FIG. 2. In calculation, we take the
same experiment parameters in Ref. [18] and Ref. [19]. Firstly, we
study the experiment [18], its slits width $a=47.5nm$, the wave
length of the ${C_{60}}$ $\lambda=2.4\times10^{-12}m$, the two
slits distance $d=52.5nm$, the distance between the slit and
screen $L=1.25m$. The theory parameters are taken:
$A_{1}=1.6\times 10^{12}$, $A_{2}=1.7\times 10^{12}$,
$c_{1}=0.915$, $c_{2}=0.40345$$(\mid c_{1}\mid^{2}+\mid
c_{2}\mid^{2}=1)$, and the quantum coherence degree $\nu=0.53$,
which is calculated by Eq. (33) with the experiment data [18].
From Eq. (33), we can obtain the calculation result, which is
shown in FIG. 3. In FIG. 3, the solid line is theoretical
calculation curve and the circle points are experiment data. we
find that the theoretical result is in accordance with the
experiment data in Ref. [18]. Finally, we study the experiment
[19], its slits width $a=42nm$, the wave length of the ${C_{60}}$
$\lambda=4.8\times10^{-12}m$, the two slits distance $d=86nm$, the
distance between the slit and screen $L=1.25m$. The theory
parameters are taken: $A_{1}=5.35\times 10^{13}$, $A_{2}=2.1\times
10^{13}$, $c_{1}=0.9075$, $c_{2}=0.42$$ ( \mid c_{1}\mid^{2}+\mid
c_{2}\mid^{2}=1 )$, and the quantum coherence degree $\nu=0.88$,
which is calculated by Eq. (33) with the experiment data [19].
From Eq. (33), we can obtain the calculation result, which is
shown in FIG. 4. In FIG. 4, the solid line is theoretical
calculation curve and the circle points are experiment data. we
find that the theoretical result is also in accordance with the
experiment data in Ref. [19].

\begin{figure}
\includegraphics[width=8.5 cm]{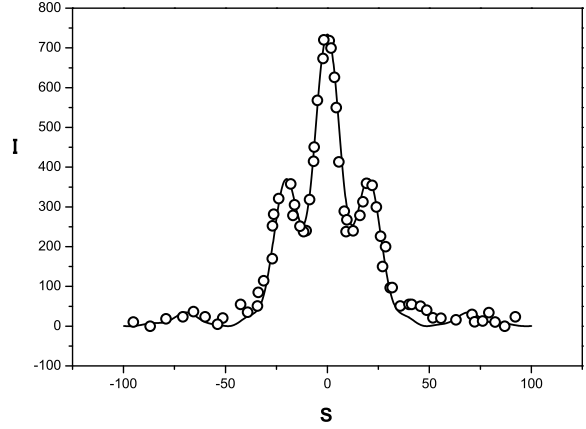}
\caption{Comparing the calculation result with the experiment data
[18].}
\end{figure}

\begin{figure}
\includegraphics[width=8.5 cm]{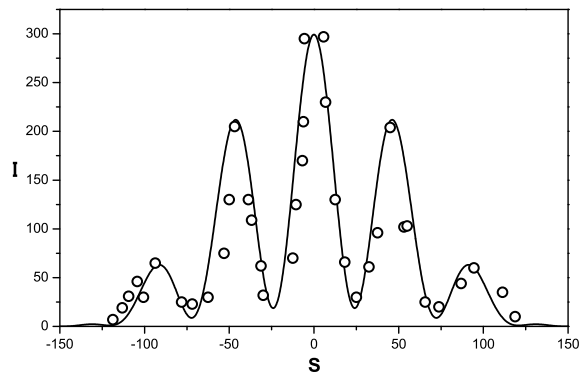}
\caption{Comparing the calculation result with the experiment data
[19].}
\end{figure}

{\bf 6. Conclusion} \vskip 8pt

In conclusion, we apply the new full quantum theory to study the
double-slit diffraction of ${C_{60}}$ molecules, and compare the
theoretical prediction with experimental data taken from Refs.
[18] and [19]. When the decoherence effects are considered, we can
find the theoretical results are in accordance with two
experimental data. This approach has universal applicability, such
as, it can also study the diffraction of electron, neutron and
atom, which include their
multi-slit and grating diffractions.\\

\end{document}